\documentclass{aa} 

\usepackage{color}

\usepackage{pgfplots}
\usepackage{tikz-3dplot}
 
\tdplotsetmaincoords{60}{115}
\pgfplotsset{compat=newest}

\usetikzlibrary{shapes.misc}
\tikzset{cross/.style={cross out, draw=black, minimum size=2*(#1-\pgflinewidth), inner sep=0pt, outer sep=0pt}, cross/.default={5pt}}

\DeclareRobustCommand{\rchi}{{\mathpalette\irchi\relax}}
\newcommand{\irchi}[2]{\raisebox{\depth}{$#1\chi$}} 

\newcommand{\ez}{\mathbf{e}_{\rm z}}

\newcommand{\rlight}{r_{\rm L}}

\usepackage{graphicx}
\usepackage{txfonts}

\usepackage{csquotes}
\usepackage{tikz}

\begin{document} 

\title{Constraining the magnetic field geometry of the millisecond pulsar PSR~J0030+0451 from joint radio, thermal X-ray and $\gamma$-ray emission}

\titlerunning{The magnetic field of PSR~J0030+0451}


\author{J. P\'etri
         \inst{1}
          \and S. Guillot \inst{2}
          \and L. Guillemot \inst{3,4}
          \and I. Cognard \inst{3,4}
          \and G. Theureau \inst{3,4,5}
          \and J.-M. Grie{\ss}meier \inst{3,4}
          \and L. Bondonneau \inst{4}
          \and D. Gonz\'alez-Caniulef \inst{2}
          \and N. Webb \inst{2}
          \and F. Jankowski \inst{3}
          \and I. P. Kravtsov \inst{9,8}
          \and J. W. McKee \inst{6,7}
          \and T. D. Carozzi \inst{10}
          \and B. Cecconi  \inst{8,4}
          \and M. Serylak \inst{11,12}
          \and P. Zarka \inst{8,4}
          }

   \institute{Universit\'e de Strasbourg, CNRS, Observatoire astronomique de Strasbourg, UMR 7550, F-67000 Strasbourg, France.\\
              \email{jerome.petri@astro.unistra.fr}   
               \and 
               IRAP, CNRS, 9 avenue du Colonel Roche, BP 44346, F-31028 Toulouse Cedex 4, France
               \and
               Laboratoire de Physique et Chimie de l'Environnement et de l'Espace, Universit\'e d'Orl\'eans / CNRS, 45071 Orl\'eans Cedex 02, France
               \and 
               Observatoire Radioastronomique de Nan\c{c}ay, Observatoire de Paris, Universit\'e PSL, Université d'Orl\'eans, CNRS, 18330 Nan\c{c}ay, France
               \and
               LUTH, Observatoire de Paris, Universit\'e PSL, Universit\'e Paris Cit\'e, CNRS, 92195 Meudon, France
               \and
                E.A. Milne Centre for Astrophysics, University of Hull, Cottingham Road, Kingston-upon-Hull, HU6 7RX, UK
               \and
               Centre of Excellence for Data Science, Artificial Intelligence and Modelling (DAIM), University of Hull, Cottingham Road, Kingston-upon-Hull, HU6 7RX, UK
               \and
               LESIA, Observatoire de Paris, Université PSL, CNRS, Sorbonne Université, Université de Paris, 5 place Jules Janssen, Meudon, 92195, France 
               \and
               Institute of Radio Astronomy of NAS of Ukraine, 4 Mystetstv St., 61002, Kharkiv, Ukraine
               \and
               Onsala Space Observatory, Chalmers University of Technology, Onsala, Sweden
               \and
               SKA Observatory, Jodrell Bank, Lower Withington, Macclesfield, SK11 9FT, United Kingdom
               \and
               Department of Physics and Astronomy, University of the Western Cape, Bellville, Cape Town, 7535, South Africa
             }

   \date{Received ; accepted }

 
  \abstract
   {With the advent of multi-wavelength electromagnetic observations of neutron stars, spanning many decades in photon energies, from radio wavelengths up to X-rays and $\gamma$-rays, it becomes possible to significantly constrain the geometry and the location of the associated emission regions.}
   {In this work, we use results from the modelling of thermal X-ray observations of PSR~J0030+0451 from the NICER mission and phase-aligned radio and $\gamma$-ray pulse profiles to constrain the geometry of an off-centred dipole able to reproduce the light-curves in these respective bands simultaneously.}
   {To this aim, we deduce a configuration with a simple dipole off-centred from the location of the centre of the thermal X-ray hot spots and show that the geometry is compatible with independent constraints from radio and $\gamma$-ray pulsations only, leading to a fixed magnetic obliquity of $\alpha \approx 75\degr$ and a line of sight inclination angle of $\zeta \approx 54\degr$.}
   {We demonstrate that an off-centred dipole cannot be rejected by accounting for the thermal X-ray pulse profiles. Moreover, the crescent shape of one spot is interpreted as the consequence of a small scale surface dipole on top of the large scale off-centred dipole. }
   {}


\maketitle

%

\section{Introduction}

Millisecond pulsars (MSPs) are old neutron stars whose spin periods ($\sim$ 1--30~ms) find their origin in the recycling scenario during which angular momentum is transferred to the neutron star via the accretion of matter from a companion star. They are also known to exhibit complex radio pulse profiles associated with strong variations in their polarization position angle (PPA) \citep{yan_polarization_2011}. This must be contrasted with young pulsars showing much more regular variation in the PPA, well described by the rotating vector model (RVM) \citep{radhakrishnan_magnetic_1969} because of the emission region occurring at higher altitude, in an almost dipole magnetic field geometry \citep{mitra_nature_2017}. The irregular behaviour of MSPs is probably related to the small size of the light cylinder and therefore to strong deviation from the dipole field and to the low altitude of emission where the multipolar components still give a contribution which is not negligible compared to the dipole component \citep{petri_illusion_2019}.  

Nevertheless, multi-wavelength observations are able to put some constraints on their magnetic configuration. For instance, a joint modelling of the radio and $\gamma$-ray pulsed emission pins down the geometry of a centred dipole for young pulsars \citep{petri_young_2021}, moreover supported by good quality radio polarization data when available. For MSPs, the situation is less constraining because the RVM cannot explain the PPA likely due to small scale magnetic field loops close to the surface, affecting the radio pulse profile and polarization angle. However, as shown by \cite{benli_constraining_2021}, phase aligned $\gamma$-ray pulse profile fitting helps to extract the magnetic obliquity~$\alpha$ of a pulsar as well as the line of sight inclination angle~$\zeta$.

As radio and $\gamma$-ray photons originate from well above the stellar surface, around several tens or hundreds of stellar radii, we do not expect these methods to allow us to disentangle the surface magnetic field structure. For example, for the pulsar studied in this paper, PSR~J0030+0451, rotating at a period of $P=4.87$~ms, the light-cylinder radius is about $\rlight \approx 232$~km corresponding to more than 19 times the neutron star radius of $R\approx12$~km. To explore the surface magnetic field, thermal X-ray pulse profile modelling looks much more promising. In the last years, the launch of the Neutron Star Interior Composition Explorer (NICER) mission led to a breakthrough in the modelling of the thermal X-ray pulsations. One of the first targets of this mission was the MSP PSR~J0030+0451, showing two prominent X-ray pulses separated by almost half a period. A careful and detailed analysis of this signal allowed to constrain the mass and radius of the neutron star \citep{riley_nicer_2019, miller_psr_2019} by fitting the thermal X-ray pulses with a model of stellar surface hot spots. This work also presented several possible shapes for the two hot spots responsible for the emission. One of the spots deviates significantly from a circular shape and resembles more to a crescent shape implying a strong non-dipolar surface magnetic field. It should be noted that other good fits were found independently by another group \citep{miller_psr_2019} implying possibly three hot spots. These spots are thought to be heated by particles accelerated to relativistic energies in the pulsar magnetosphere before impacting the region where the magnetic field lines meet the neutron star surface \citep[][see also \citealt{baubock_atmospheric_2019, salmi_magnetospheric_2020}]{sznajder_heating_2020} and should therefore be related to the magnetic field structure and plasma flow within the neutron star magnetosphere. Before the NICER era, \cite{ruderman_neutron_1991} and more recently \cite{bogdanov_thermal_2008} already suggested the presence of an off-centred dipole to fit the X-ray pulse profile. However this seems insufficient to fully account for the recent NICER results because some quadrupolar components seemed to be also required. A comparison between the expectations from the NICER observations about the geometry of the magnetic moment and the inclination of the line of sight with older works like \cite{johnson_constraints_2014} is discussed in \cite{bilous_nicer_2019}. In the outer gap model of \cite{johnson_constraints_2014} the line of sight inclination angle was found to be about $\zeta=68\degr \pm 1\degr$ whereas \cite{riley_nicer_2019} found $\zeta=53.9\degr\,(+6.3\degr,-5.7\degr)$ for PSR~J0030+0451.

From a theoretical point of view, several groups attempted to connect these observations to their results about neutron star magnetosphere simulations. For instance, \cite{kalapotharakos_multipolar_2021} used an off-centred dipole+quadrupole field structure to reproduce the polar cap shapes as well as the radio and $\gamma$-ray light curves. Following the same idea, \cite{carrasco_relativistic_2023} performed general-relativistic force-free simulations of a centred dipole magnetic field only, introducing non standard emission regions based on the current density hitting the surface in order to model the thermal X-ray radiation for four NICER MSPs (PSR J0437$-$4715, PSR J1231$-$1411, PSR J2124$-$3358, and PSR J0030+0451). In this model, no off-centred dipole or multipoles are required. This contrasts also with \cite{chen_numerical_2020} who modelled a magnetic field geometry including an off-centred dipole+quadrupole to reproduce the polar cap shape found by the NICER collaboration. However in order to produce efficiently pair cascading close to the surface, it is known that small curvature radius are required, one to two orders of magnitude smaller than for the dipole. These smaller radii can be achieved by adding a small scale dipole anchored in the neutron star crust \citep{gil_modelling_2002}. Solving for the magnetic field topology is certainly one of the most difficult and central problems in neutron star physics remaining to be solved \citep{petri_illusion_2019}. 

In this paper we show that a large scale off-centred magnetic dipole configuration is compatible with PSR~J0030+0451 multi-wavelength observations. We show that the joint radio and $\gamma$-ray pulse profile modelling leads independently to the same conclusion as the thermal X-ray pulse profile fitting. The dominant magnetic field structure is consistent with an off-centred dipole and the peculiar hot spot crescent shape can be attributed to a small scale dipole localized in the vicinity of one pole. Section~\ref{sec:Donnees} summarizes the multi-wavelength data used in the present study and the analysis technique. Section~\ref{sec:Field} describes the method to deduce the magnetic field structure from the hot spot location. A discussion of the results is given in Section~\ref{sec:Discussion} before concluding in Section~\ref{sec:Conclusions}.


\section{Multi-wavelength pulse profile data sets}
\label{sec:Donnees}

\subsection{Radio pulse profiles}
\label{sec:radio_data}

\subsubsection{NRT}

The Nan\c{c}ay Radio Telescope (NRT) is a meridian telescope equivalent to a 94-m parabolic dish located at the Nan\c{c}ay Radio Observatory near Orl\'eans, France. Since the early 2000s, PSR~J0030+0451 has been routinely observed with the NRT for timing purposes, with an average cadence of about 10 observations per month and with the bulk of the observations conducted at a central frequency near 1.4~GHz. 

As mentioned in the subsequent Section~\ref{sec:gamma_data}, the $\gamma$-ray pulse profile for PSR~J0030+0451 used in this work was taken from the Second \textit{Fermi} Large Area Telescope Catalog of $\gamma$-ray Pulsars, hereafter 2PC \citep{abdo_second_2013}. For the $\gamma$-ray analysis of PSR~J0030+0451, a pulsar timing solution for the pulsar constructed using NRT data was used. The timing solution was built by analysing data recorded with the Berkeley-Orl\'eans-Nan\c{c}ay (BON) pulsar backend, between October~2004 and August~2011. The corresponding 1.4~GHz BON pulse profile shown in Fig.~\ref{fig:j00300451multilambda} and the timing solution can be retrieved from the 2PC archive of auxiliary files\footnote{See \url{https://fermi.gsfc.nasa.gov/ssc/data/access/lat/2nd_PSR_catalog/}}. 

The NICER X-ray data analysed as part of this work were recorded more recently, as further described in Section~\ref{sec:xray_data}. We therefore used NRT data to build another timing solution for PSR~J0030+0451, contemporaneous with the NICER data. The radio data considered in this new analysis were recorded with the NUPPI backend (a version of the Green Bank Ultimate Pulsar Processing Instrument designed for the NRT), which replaced the BON backend as the principal pulsar timing instrument in operation at Nan\c{c}ay in August~2011. We used pulsar timing data recorded between August~2011 and April~2022. Although the two pulsar timing solutions described in this Section were obtained by analysing data from the same telescope, the two timing solutions were obtained from two separate analyses; therefore, the two ephemerides did not necessarily assume a common reference phase for the radio pulse profiles. We compared the BON and NUPPI reference profiles to determine the offset between the two reference phases. This phase offset of $\sim 0.056$ was added to the NICER photon phases, in order for the radio, X-ray and $\gamma$-ray pulse profiles to share a common phase reference. 


\subsubsection{LOFAR}
We also observed J0030+0451 using the international LOFAR station SE607 in Onsala, Sweden. 
LOFAR is fully described in \citet{stappers_observing_2011} and \citet{van_haarlem_lofar_2013}.
LOFAR stations have two different frequency bands: We used the HBA
band, for which the station consists of 96 antenna tiles. Each of these tiles is made up of 16 dual-polarization antenna elements.
For each tile, an analog sum is performed. The tile signals are then digitized, and added numerically to form a coherent beam.
For this project, we recorded data between 110.0 to 187.9 MHz.
We re-folded the LOFAR observations using the NUPPI timing solution, except for the dispersion measure, for which we directly calculated one DM value per LOFAR observation.
We combined 168 observations recorded between 2019 and 2023, correcting for time-dependent DM variations.

\subsubsection{NenuFAR}

We also observed J0030+0451 using the new low frequency radio telescope NenuFAR \citep[New extension in Nan\c cay upgrading LOFAR, see][]{zarka_nenufar_2015, zarka_low-frequency_2020} 
NenuFAR is a compact phased array and interferometer formed of hexagonal groups of 19 dual-polarization antennas called mini-arrays. It is located in Nan\c{c}ay, and operates in the 10--85~MHz frequency range . We used the dedicated pulsar instrumentation UnDysPuTeD and associated software LUPPI. LUPPI is based on NUPPI, and is designed to dedisperse and fold pulsar observations in real time, with up to four bands of 37.5~MHz simultaneously \citep{bondonneau_pulsars_2021}. 
NenuFAR is an array of antennas, rolled out in several construction phases since the start of its observations in 2019. Observations in 2019 -- 2022 used 56 mini-arrays and observations in 2022 -- 2023 used 80 mini-arrays, out of which a minimum of 75 are active at any point, i.e.\ $\geq$1425 antennas in total, allowing for sensitive observations \citep{petri_constraining_2023}.
We recorded data between 20.3 and 84.3 MHz; in the final analysis, we only kept data above 42 MHz as scattering started to considerably affect the profile shape at the lowest frequencies.
We re-folded the NenuFAR observations using the NUPPI timing solution, except for the DM value, for which the value determined directly from the NenuFAR observations is considerably more precise owing to the low-frequency coverage of NenuFAR. 
We combined 95 observations, correcting for time-dependent DM variations. We then added a constant phase offset to the NenuFAR observations to correct for different instrumental delays between NenuFAR and NUPPI.   

J0030+0451 is among the twelve MSPs detected by NenuFAR, see \citet{bondonneau_pulsars_2021} and Bondonneau et al. (in prep.). The NenuFAR profile of the pulsar differs from that obtained by NRT, but it shows a comparable width.
Furthermore, our observations are compatible with the radio emission detected by NenuFAR and NRT being emitted at the same rotational phase of the pulsar.

\subsection{Thermal X-ray pulse profile with NICER data}
\label{sec:xray_data}

The thermal X-ray pulse profile was obtained from NICER observations of PSR~J0030+0451. We used all available data from 2017-07-24 20:36:00 to 2022-02-12 15:10:00, corresponding to ObsIDs 1060020101 to 4060020619.  To process the data, we employed the standard recipes with nicerl2 (from \texttt{NICERDAS v9} distributed with \texttt{HEASOFT 6.30} and NICER calibration file v20210707). Furthermore, we used the \texttt{NICERsoft}\footnote{\url{https://github.com/paulray/NICERsoft}} suite to perform additional filtering, with the following criteria beyond the default ones. We excluded the detector \# 34, known to be particularly noisy; we excluded portions of the NICER orbit during which the Earth magnetic cut-off rigidity was lower than 1.5 GeV/c at the satellite's location. We also restricted the data to observing time when the space weather index KP was lower than 5, and when undershoot were lower than 200~c/s (see the NICER Data analysis threads \footnote{\url{https://heasarc.gsfc.nasa.gov/docs/nicer/analysis_threads/}} for detailed description of these filtering criteria).

The \texttt{NICERsoft} package was also used to merge the event files into a single event file. Finally, we also performed a count rate cut on the merged event file, first when detector \#14 (also known to be noisy on occasion) had a count rate above 1~c/s (with 8-sec time bins), and second when the total count rate (all remaining detectors) was larger than 6~c/s to remove generally noisy time intervals remaining after the filtering described above. The phases of all events were calculated with the \texttt{photonphase} task of the \texttt{PINT} package \footnote{\url{https://nanograv-pint.readthedocs.io}}, and using the NUPPI ephemeris, including a phase offset of $\sim 0.056$, as described in Section~\ref{sec:radio_data} 

\begin{figure}
	\centering
	\includegraphics[width=\linewidth]{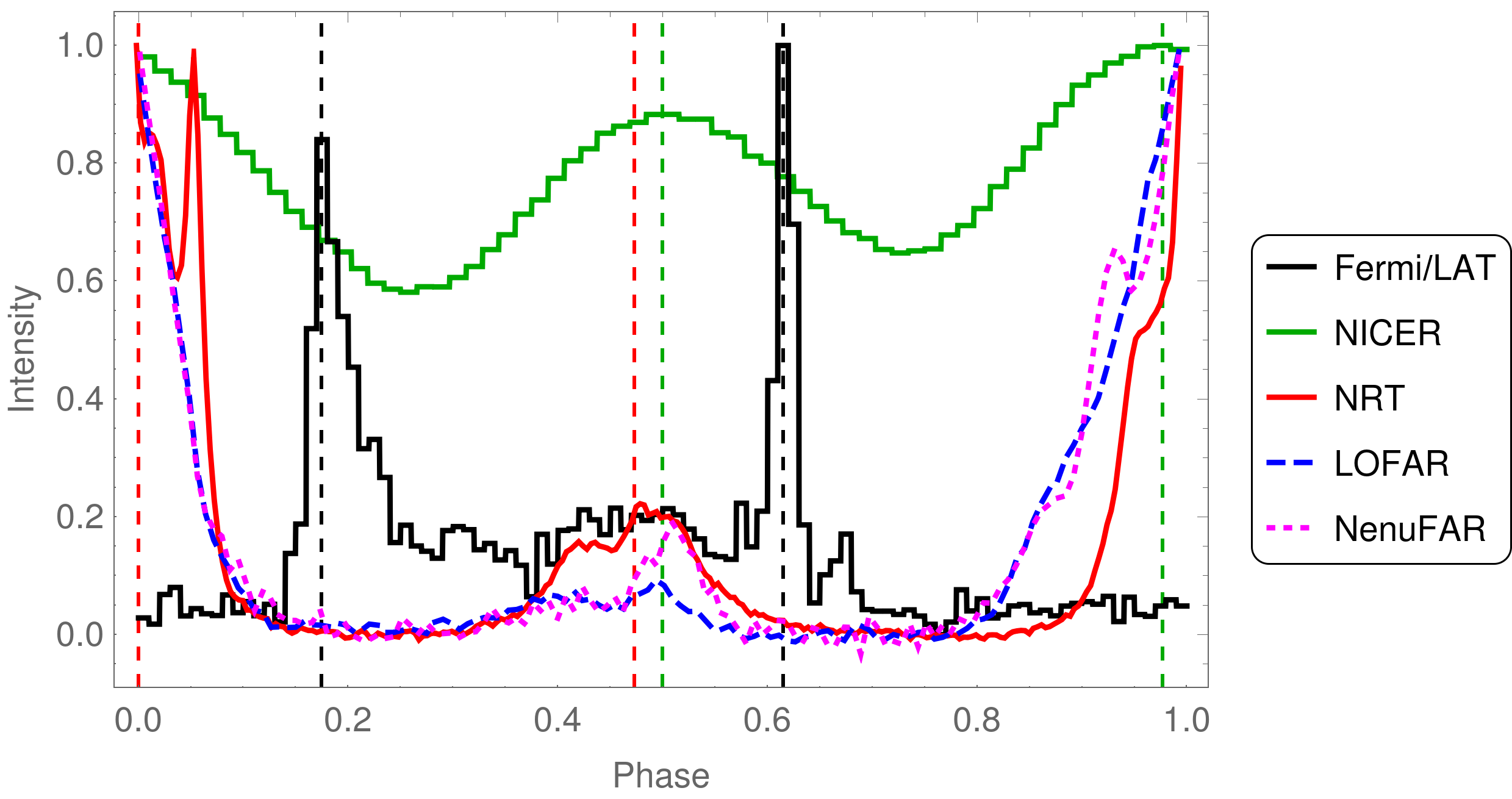}
	\caption{Time lags between peaks in the different wavelengths. Radio pulse profiles are in red for NRT, in dashed blue for LOFAR, in dotted magenta for NenuFAR, thermal X-ray are in green, and Fermi/LAT are in black. The dashed coloured vertical lines show the phase of the peak flux of each pulse.}
	\label{fig:j00300451multilambda}
\end{figure}

Given the relative faintness of thermal X-ray emission of MSPs compared to the background level of NICER observations, we also calculated the optimal energy range that maximises the detection of the pulsations (see \cite{guillot_nicer_2019} for the description of this optimisation).  This energy range will vary from pulsar to pulsar, and depends on the spectral shape of the pulsar, the energy range where the NICER effective area is the largest, and on the overall background spectral shape of the observations (which can vary depending on the strength of the various background components). For PSR~J0030+0451, we find an optimal energy range of 0.28--1.46 keV, giving a detection significance of 223~sigma. The pulse profile is shown in Fig.~\ref{fig:j00300451multilambda}.

\subsection{$\gamma$-ray pulse profile}
\label{sec:gamma_data}

Similar to the BON radio pulse profile for PSR~J0030+0451 displayed in Fig.~\ref{fig:j00300451multilambda}, the $\gamma$-ray pulse profile was taken from the 2PC auxiliary files archive. The profile, also displayed in Fig.~\ref{fig:j00300451multilambda}, was constructed as part of the preparation of 2PC by analysing $\gamma$-ray photons recorded by the \textit{Fermi} Large Area Telescope (LAT) between 2008 August~4 and 2011 August~4. $\gamma$-ray photons with reconstructed energies from 0.1 to 100~GeV were selected and phase-folded with the BON ephemeris for PSR~J0030+0451 as described in Section~\ref{sec:radio_data}.


\section{Pulsar magnetic field determination}
\label{sec:Field}

Based on the radio and $\gamma$-ray data, the geometry of the dipole magnetic field of PSR~J0030+0451 has already been investigated with the striped wind model in \cite{petri_unified_2011} relying on the split monopole configuration introduced by \cite{bogovalov_physics_1999}. More recently \cite{benli_constraining_2021} used the force-free magnetosphere solution for the special class of millisecond pulsars. In the latter work, the best fit parameters were found to be around $\alpha=70\degr$ and $\zeta=60\degr$ where $\alpha$ is the magnetic obliquity and $\zeta$ the line-of-sight inclination angle relative to the spin axis. In the next section, we re-explore these results, showing several possible configurations for the joint radio and $\gamma$-ray light curve fitting. Then we constrain the off-centred position of the dipole from the thermal X-ray hot spot location and shape.

\subsection{Dipole geometry from joint radio and $\gamma$-rays modelling}

The light-curve fitting procedure has been explained in depth in \cite{benli_constraining_2021} as well as in \cite{petri_young_2021}. We simply recall that the two main characteristics to be adjusted are the $\gamma$-ray peak separation~$\Delta$ and the phase lag between the first $\gamma$-ray peak and the peak of the radio profile~$\delta$. To a very good approximation, with an error of less than 1\%, \cite{petri_unified_2011} showed that this relation is
\begin{equation}\label{eq:cos_pi_delta}
\cos (\pi \, \Delta) = |\cot \alpha \cot \zeta| .
\end{equation}
For PSR~J0030+0451, the second Fermi pulsar catalogue \citep{abdo_second_2013} reports a value of the phase separation between the two $\gamma$-ray peaks of $\Delta = 0.450$ as well as a $\gamma$-ray lag of the first $\gamma$-ray peak with respect to the peak of the radio profile of $\delta = 0.146$. Additionally the NICER collaboration reports a value of $\zeta=54\degr$ \citep{riley_nicer_2019}. This would imply an obliquity of $\alpha=78\degr$ not far from the independent $\gamma$-ray light curve fitting done by \cite{benli_constraining_2021} who found $\alpha=70\degr$. Using the input from the thermal X-ray detailed in Section~\ref{sec:Xray}, we were also looking for solutions of the $\gamma$-ray light-curves by constraining the line of sight inclination to $\zeta=54\degr$ or to $\zeta=50\degr$ as given by the NICER collaboration.

\begin{figure}
	\centering
	\includegraphics[width=\linewidth]{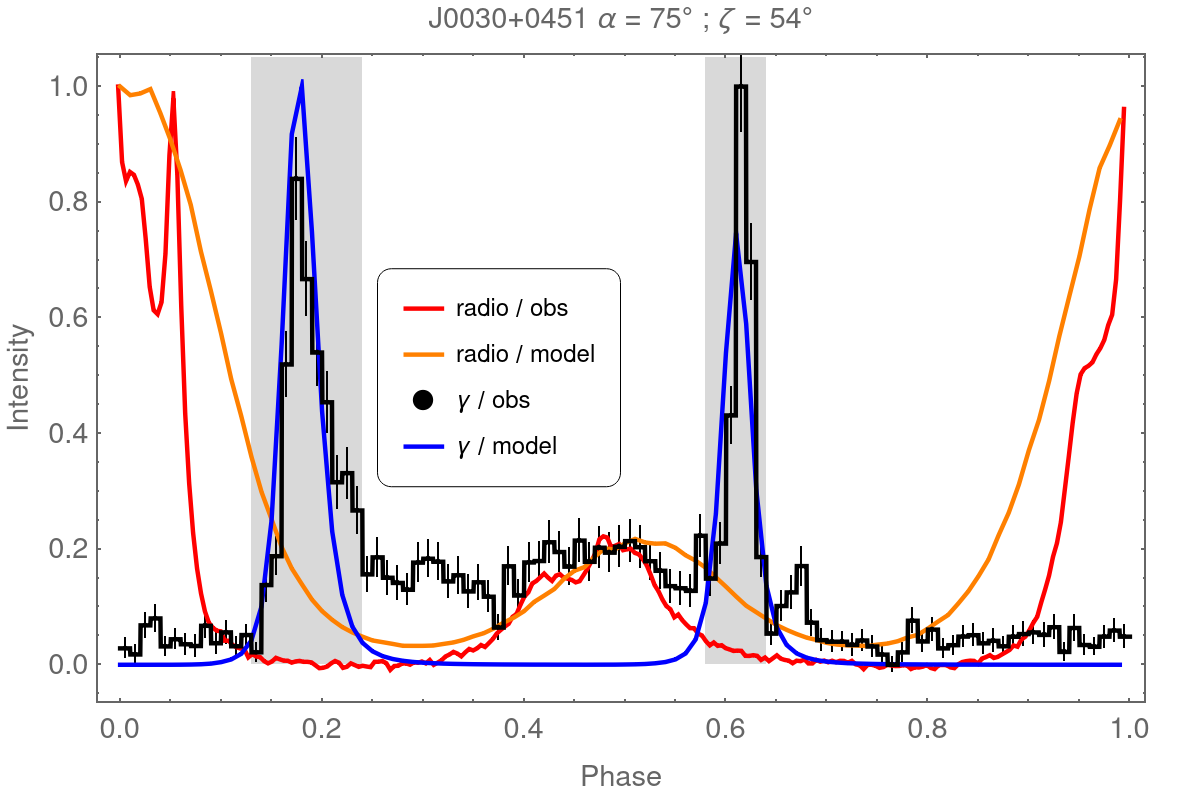} \\
	\includegraphics[width=\linewidth]{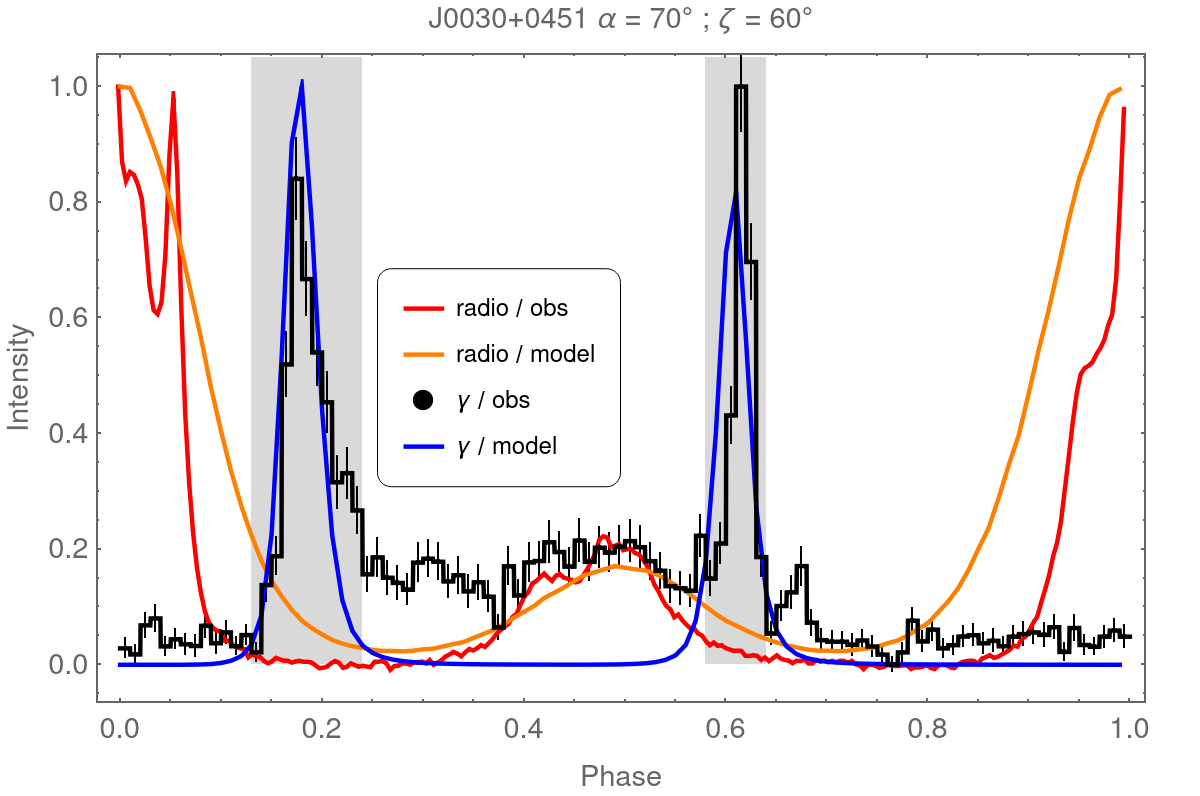}
	\caption{Some fits for the radio and $\gamma$-ray light curves of PSR~J0030+0451. The parameters used are $\alpha=75\degr$ and $\zeta=54\degr$ on the top panel and $\alpha=70\degr$ and $\zeta=60\degr$ on the bottom panel. The gray coloured boxes show the phase intervals used for the $\gamma$-ray fit.}
	\label{fig:fitj00300451}
\end{figure}

To find a good fit to the data, we minimize a kind of reduced $\rchi^2_\nu$ value defined by
\begin{equation}
    \rchi^2_\nu = \frac{1}{\nu} \sum_{i=1}^N \left( \frac{I_i^{\rm mod}-I_i^{\rm obs}}{\sigma_i} \right)^2,
\end{equation}
\noindent where $N$ is the number of data points, $\nu=N-2$ the degree of freedom, $I_i^{\rm obs}$ the observed $\gamma$-ray flux and $I_i^{\rm mod}$ the predicted $\gamma$-ray flux. The number two parameters to adjust, the magnetic obliquity and the inclination of the line of sight. The most important characteristics of our model is the $\gamma$-ray peak separation~$\Delta$ and the $\gamma$-ray lag~$\delta$. Our model does not produce off-pulse emission. Therefore we select only the phase intervals containing the two peaks, as shown in the gray coloured boxes in Fig.~\ref{fig:fitj00300451}. All light-curves are plotted in arbitrary units, applying a normalization such that the peak value in each energy band is unity. The best values for the parameters $\alpha$, $\zeta$ and the shift in phase $\phi_s$ are those minimizing the $\rchi^2_\nu$. Two possible fits are shown in Fig.~\ref{fig:fitj00300451} and agree reasonably well between radio, $\gamma$-rays and the geometry constrained from thermal X-rays. We found $\alpha=75\degr$ and $\zeta=54\degr$ for the upper panel and $\alpha=70\degr$ and $\zeta=60\degr$ for the lower panel. These results rely on a centred dipole, thus assuming that the radio and $\gamma$-ray emission emanate from high altitude, where the perturbation of surface fields and the off-centering of the large scale dipole become imperceptible due to the faster decrease of these components with distance. We next show how to determine the close region field structure, where the off-centred position of the large scale dipole is crucial for the thermal surface emission.

\subsection{Off-centred dipole geometry from thermal X-rays}
\label{sec:Xray}

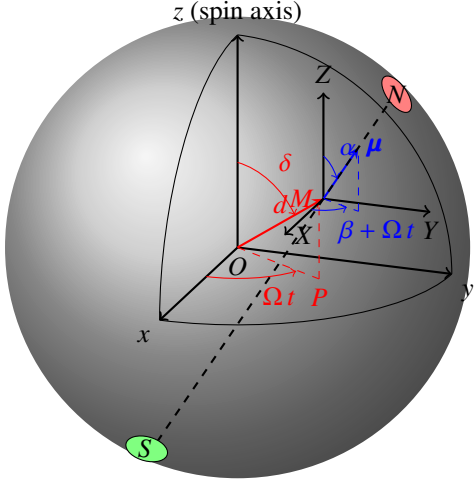
\begin{figure}
	\centering

\tdplotsetmaincoords{70}{110}

\pgfmathsetmacro{\rvec}{.6}
\pgfmathsetmacro{\thetavec}{55}
\pgfmathsetmacro{\phivec}{70}

\begin{tikzpicture}[scale=3,tdplot_main_coords]

\coordinate (O) at (0,0,0);

\tdplotsetcoord{P}{\rvec}{\thetavec}{\phivec}

\tdplotsetcoord{A}{1}{60}{0}

\tdplotsetcoord{B}{1.5}{60}{0}


\shade[ball color = lightgray,
opacity = 0.2
] (0,0,0) circle (1.02cm);

\draw[thick,->] (0,0,0) node[below] {$O$} -- (1,0,0) node[anchor=north east]{$x$};
\draw[thick,->] (0,0,0) -- (0,1,0) node[anchor=north west]{$y$};
\draw[thick,->] (0,0,0) -- (0,0,1) node[anchor=south]{$z$ (spin axis)};

\tdplotsetrotatedcoords{0}{0}{0}

\draw[-stealth,thick,color=red] (O) -- (P) node [midway, above] {$d$} node [left] {$M$} ;
\draw[dashed,color=red,tdplot_rotated_coords] (0,0,0) -- (0.26,0.475,0) node [below] {$P$} ;
\draw[dashed,color=red,tdplot_rotated_coords] (0.26,0.475,0) -- (0.26,0.475,.35) ;
\tdplotdrawarc[tdplot_rotated_coords,color=red,->]{(0,0,0)}{0.4}{0}{60}{anchor=north west}{$\Omega\,t$}




\tdplotsetthetaplanecoords{\phivec}

\tdplotdrawarc[red, tdplot_rotated_coords,->]{(0,0,0)}{0.4}{0}{\thetavec}{anchor=south west}{$\delta$}

\tdplotsetthetaplanecoords{0}



\tdplotsetrotatedcoords{0}{0}{0}

\tdplotsetrotatedcoordsorigin{(P)}

\draw[thick,tdplot_rotated_coords,->] (0,0,0) -- (.5,0,0) node[anchor=west]{$X$};
\draw[thick,tdplot_rotated_coords,->] (0,0,0) -- (0,.5,0) node[anchor=north]{$Y$};
\draw[thick,tdplot_rotated_coords,->] (0,0,0) -- (0,0,.5) node[anchor=south]{$Z$};


\draw[-stealth,thick,color=blue,tdplot_rotated_coords] (0,0,0) -- (.1,.2,.3) node [right] {$\pmb{\mu}$} ;
\draw[dashed,color=blue,tdplot_rotated_coords] (0,0,0) -- (.1,.2,0);
\draw[dashed,color=blue,tdplot_rotated_coords] (.1,.2,0) -- (.1,.2,.3);
\draw[dashed,thick,color=black,tdplot_rotated_coords] (-.5,-1,-1.5) -- (.2,.4,.6) ;
\tdplotdrawarc[tdplot_rotated_coords,color=blue,->]{(0,0,0)}{0.15}{0}{63}{anchor=north west}{$\beta+\Omega\,t$}

\tdplotsetrotatedthetaplanecoords{45}

\tdplotdrawarc[tdplot_rotated_coords,color=blue,->]{(0,0,0)}{0.2}{0}{45}{anchor=south west}{$\alpha$}

\begin{scope}[canvas is xy plane at z=0]
\draw (1,0) arc (0:90:1);
\end{scope}
\begin{scope}[canvas is xz plane at y=0]
\draw (1,0) arc (0:90:1);
\end{scope}
\begin{scope}[canvas is yz plane at x=0]
\draw (1,0) arc (0:90:1);
\end{scope}

\coordinate (N) at ({0/sqrt(3)},{1.05/sqrt(2)},{1.15/sqrt(2)});
\coordinate (S) at ({0/sqrt(2)},{-0.6/sqrt(2)},{-1/sqrt(1)});
\coordinate (Q) at ({0.19245, -0.278954, 0.663855}) ;

\pgfmathsetmacro{\r}{.3} %
\pgfmathsetmacro{\h}{0.95*\r} %
\tdplotsetrotatedcoords{90}{-130}{10}%
\draw[black, fill = red!50, tdplot_rotated_coords] (N) node{$N$}  circle[radius=sqrt(\r*\r-\h*\h)];
\tdplotsetrotatedcoords{60}{-160}{0}%
\draw[black, fill = green!50, tdplot_rotated_coords] (S) node{$S$}  circle[radius=sqrt(\r*\r-\h*\h)];
\end{tikzpicture}

    \caption{Geometry of the off-centred dipole showing the location of the hot spots $N$ and $S$ and the angles defined in the text $(\alpha, \delta, \beta)$.}
    \label{fig:hotspot}
\end{figure}
We constrain the geometry of the off-centred large scale dipole from both the location and size of the two polar caps that are deduced from the NICER observations of \cite{riley_nicer_2019} and \cite{miller_psr_2019} and the half-opening angles of the cones subtending their rims. Fig.~\ref{fig:hotspot} highlights the geometry and shows the relevant angles and axes. Let us assume that the centre of the north polar cap is located at the point~$N$ with spherical coordinates $(R,\theta_{\rm n},\phi_{\rm n})$ and the position vector given by $\vec{R}_{\rm n} = R\,\vec{n}$ and similarly for the south polar cap at point~$S$ with coordinates $(R,\theta_{\rm s},\phi_{\rm s})$ and a position vector $\vec{R}_{\rm s} = R\,\vec{s}$ where $R$ is the neutron star radius. $\vec{n}$ and $\vec{s}$ are therefore unit vectors pointing towards the centre of the hot spots. Let us also denote the unit vector joining both polar cap centres by
\begin{equation}
 \vec{t} = \frac{\vec{n}-\vec{s}}{\|\vec{n}-\vec{s}\|}.
\end{equation}
The obliquity of the magnetic dipole is therefore given by projection onto the rotation axis $\ez$ according to
\begin{equation}
\cos \alpha = \pm \vec{t} \cdot \ez .
\end{equation}
We choose the solution corresponding to $\alpha \in [0, \pi/2]$.
The minimal distance of the line joining $N$ and $S$ to the centre of the neutron star is given by
\begin{equation}\label{eq:epsilon}
 \epsilon = \frac{d}{R} = \sqrt{1 - (\vec{s} \cdot \vec{t})^2 } .
\end{equation}
This point on the segment $NS$ is denoted by $M$.
Equation~\eqref{eq:epsilon} gives a lower bound to the distance~$d$ from the magnetic dipole moment location to the star centre. Would the dipole be located at any other point along the segment joining the point~$N$ to the point~$S$, the distance~$d$ between the dipole position~$M$ and the centre of the star~$O$ would be larger.
The angle between the line joining the dipole moment to the star centre and the rotation axis is
\begin{equation}
\cos \delta = \vec{m} \cdot \ez
\end{equation}
where $\vec{m} = \overrightarrow{OM}/OM$ and $\overrightarrow{OM} = R \, (\vec{s} - (\vec{s} \cdot \vec{t}) \, \vec{t})$.
The angle~$\beta$ is the angle between the vector $\overrightarrow{OM}$ projected onto the plane $(xOy)$ written as $\overrightarrow{OP}$ and the vector joining $S$ and $N$ thus $\vec{n}-\vec{s}$ or the unit vector $\vec{t}$
\begin{equation}
\cos \beta = \overrightarrow{OP} \cdot \vec{t} / OP.
\end{equation}
The relative angular size difference between the north and the south hot spot region helps to constrain the location of the magnetic dipole along the line joining the two poles $N$ and $S$. Indeed, knowing the north and south polar cap radius respectively denoted as $\xi_{\rm n}$ and $\xi_{\rm s}$, their ratio~$\xi_{\rm n}/\xi_{\rm s}$ is related to the distance~$d$ between the location of the dipole moment and the centre of the star. 
For instance, for an aligned rotator, if the dipole is located at the stellar centre, this ratio is equal to unity because both spots are identical. Moreover, using a spherical coordinate system $(r,\theta,\phi)$, the field lines are given by $r = \lambda \, \rlight \sin^2\theta$ with $\rlight=c/\Omega$ the light-cylinder radius, $c$ the speed of light, $\Omega$ the stellar spin frequency and $\lambda$ a parameter labelling each field line. 
However, already when the dipole is shifted along the rotation axis, the geometry becomes asymmetrical and one spot grows whereas the second shrinks. More quantitatively, the half-opening angle $\xi$ (north or south spot) is solution to the transcendental equation 
\begin{multline}\label{eq:Rapport}
a^2 = \lambda^2 \, \sin^4\xi + \epsilon^2 \, a^2 - 2 \, \lambda \, \epsilon \, a \, \sin^2\xi \\
 (\cos\xi \, \cos \theta_0 + \sin\xi \, \sin\theta_0 \, \cos(\phi-\phi_0)) ,
\end{multline}
with $a=R/\rlight$ and
\begin{equation}
\lambda = \epsilon \, a \, \sin\theta_0 \, \cos(\phi-\phi_0) \pm \sqrt{1 - \epsilon^2 \, a^2 \, \sin^2\theta_0 \, \sin^2(\phi-\phi_0) } ,
\end{equation}
$\theta_0$ and $\phi_0$ the spherical polar colatitude and azimuth of the magnetic moment position. For an arbitrary location of this magnetic moment, even for an aligned rotator, the polar cap shapes deviate from a circular rim and the axial symmetry is broken. 
Therefore we need to solve for the radius at each azimuth~$\phi$ to deduce the opening angle~$\xi$ for given $\epsilon$ and $a$. We need to find the roots $\xi(\phi)$ of Eq.\eqref{eq:Rapport}. 
The sign of $\lambda$ is arbitrary because the solution always shows a symmetry between the north pole and the south pole. For a centred dipole $\epsilon=0$ and $\lambda=1$, thus $\sin\xi = \sqrt{a} = \sqrt{R/\rlight}$ as is well known. A similar computation could be done with an orthogonal static rotator. The polar cap shape is then given by \cite{petri_general-relativistic_2018} and depends on the angle $\phi$. We do not enter into such complications because accurate determination of the rims would require a full force-free numerical simulation of the magnetosphere which is too computationally expensive in regard to the number of free parameters to explore. However, the aligned case already gives a good insight into the impact of off-centring onto the polar cap respective sizes.

\begin{figure}
	\centering
	\includegraphics[width=0.9\linewidth]{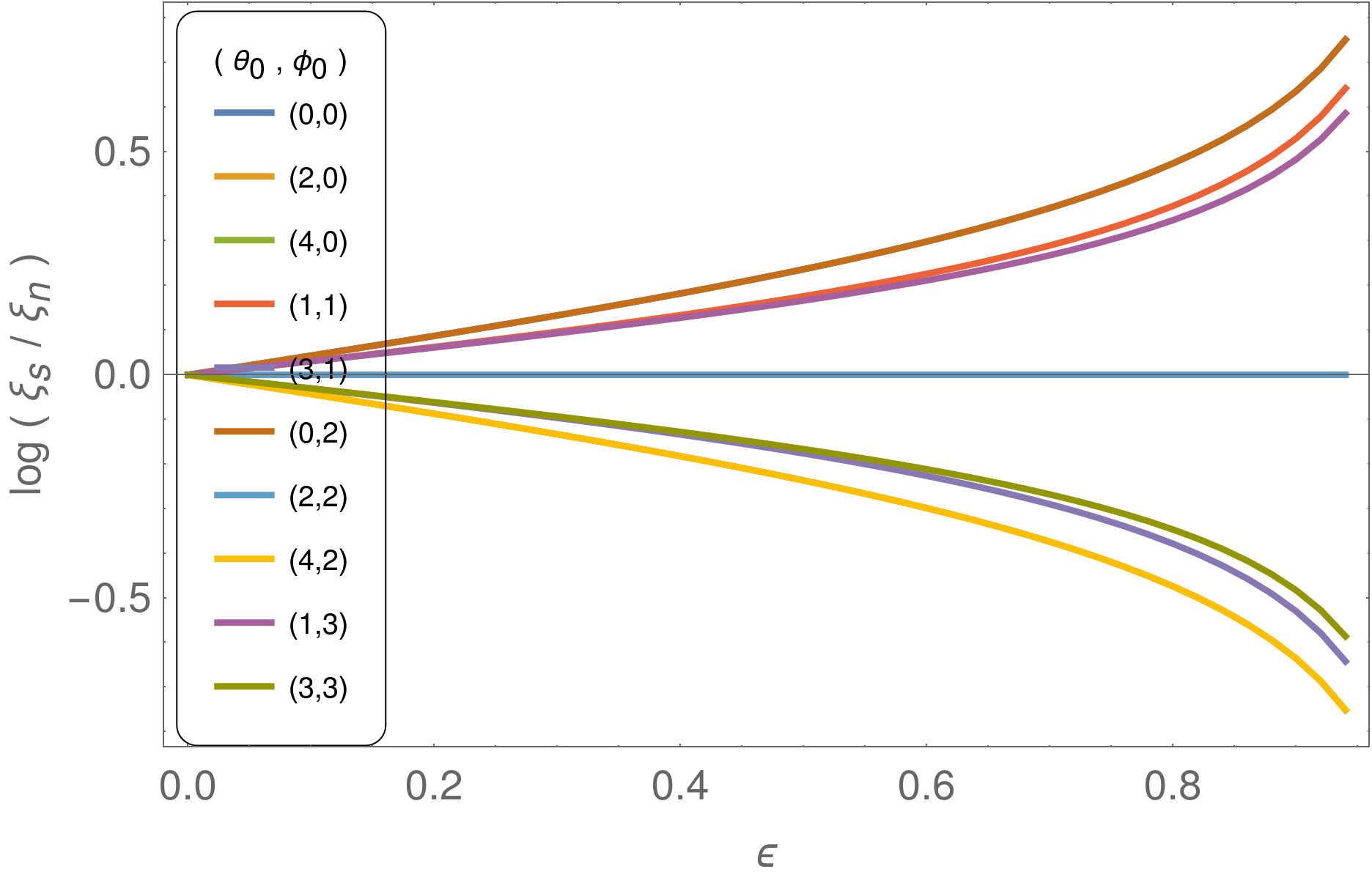}
	\caption{Ratio $\xi_{\rm s}/\xi_{\rm n}$ of the polar cap half-opening angles of the south and north pole respectively $\xi_{\rm s}$ and $\xi_{\rm n}$ as a function of the fractional offset from the centre $\epsilon=d/R$. The colours correspond to different values of the angles $(\theta_0,\phi_0)$ in units of $\pi/4$.}
	\label{fig:ratioanglecalotte}
\end{figure}

In the particular case of the millisecond pulsar PSR~J0030+0451, the ratio of stellar radius to light-cylinder radius is $a=R/\rlight \approx 0.055$ according to \cite{riley_nicer_2019} who found an estimate of its radius $R=12.71^{+1.14}_{-1.19}$~km and its mass $M=1.34^{+0.15}_{-0.16}\ M_\odot$. 
The offset value $\epsilon=d/R$ required to adjust the relative spot angular sizes for different values of $\alpha$ and $\phi$ according to Eq.\eqref{eq:Rapport}
is shown in Fig.~\ref{fig:ratioanglecalotte}. $\theta_0$ and $\phi_0$ in the legend are given in units of $\pi/4$ for several different values spanning the range $[0,\pi]$ in steps of $k\,\pi/4$ with $k\in[0..4]$. Because some curves overlap, they are not all shown. A ratio $\xi_{\rm s}/\xi_{\rm n}$ larger that 4 requires an offset as high as $\epsilon\gtrsim0.9$. A ratio $\xi_{\rm s}/\xi_{\rm n}$ equal to unity means that the two spots are of equal size. In Fig.~\ref{fig:ratioanglecalotte} we show the ratio $\xi_{\rm s}/\xi_{\rm n}$ in log scale.  
We notice that this plot is symmetric with respect to the x-axis. Translated into the ratio $\xi_{\rm s}/\xi_{\rm n}$, it means that for each value of displacement~$\epsilon$ there exist a value~$r_1$ of the ratio $\xi_{\rm s}/\xi_{\rm n}=r_1$ with a position of the magnetic moment at $(\theta_1,\phi_1)$ associated to the inverse ratio $\xi_{\rm s}/\xi_{\rm n} = 1/r_1$ with a new position of the magnetic moment given by the angles $(\theta_2,\phi_2)$ uniquely related to $(\theta_1,\phi_1)$. Such symmetry is expected because of the symmetric nature of the magnetic dipole with respect to the inversion between north pole and south pole. For a more realistic force-free field these values could differ but we expect also large offsets for large differences in the spots sizes.

The fit parameters of \cite{riley_nicer_2019} are summarized in Table~\ref{tab:NICER1} that shows some of the models they used. Among others, these are the single temperature+protruding single temperature (ST+PST), the single temperature+eccentric single temperature (ST+EST), the single temperature+concentric single temperature (ST+CST), the single temperature unshared parameters (ST-U), the concentric dual-temperature unshared parameters (CDT-U).
The geometry of the off-centred dipole with $(\alpha, \beta, \delta, \epsilon)$ is then deduced and shown on the right columns in the same table.
\begin{table*}
	\centering
\begin{tabular}{lcccccccccccc}
	\hline
	Hot spot & $\Theta_p$ & $\phi_p$ & $\xi_p$ & $\Theta_s$ & $\phi_s$ & $\xi_s$ & $\xi_s/\xi_p$ & $\zeta$ & $\alpha$ & $\beta$ & $\delta$ & $\epsilon$ \\
	Model & (rad) & (cycle) & (rad) & (rad) & (cycle) & (rad) &  & (deg) & (deg) & (deg) & (deg) & ($d/R$) \\
	\hline \hline
	ST+PST & 2.23 & 0.46 & 0.09 & 2.91 & $-$0.59 & 0.46 & 5.1 & 54 & 70 & 22 & 160 & 0.84 \\
	ST+EST & 2.22 & 0.45 & 0.07 & 2.66 & $-$0.51 & 0.28 & 4.0 & 58 & 77 & 26 & 166 & 0.77 \\
	ST+CST & 2.24 & 0.46 & 0.07 & 2.60 & $-$0.50 & 0.27 & 3.9 & 58 & 79 & 31 & 168 & 0.75 \\
	ST-U   & 2.48 & 0.46 & 0.14 & 2.78 & $-$0.50 & 0.29 & 2.1 & 60 & 81 & 24 & 170 & 0.87 \\
	CDT-U  & 2.24 & 0.46 & 0.15 & 2.61 & $-$0.50 & 0.27 & 1.8 & 58 & 79 & 30 & 167 & 0.76 \\
	\hline
\end{tabular}
\caption{Off-centred dipole geometry deduced from the polar cap location after \cite{riley_nicer_2019}.\label{tab:NICER1}
}
\end{table*}

Independent constraints from a second group \citep{miller_psr_2019} are reported in Table~\ref{tab:NICER2} for a 2~spot and a 3~spot model\footnote{As noted in  \cite{miller_psr_2019}, however, the 3-spot model is not statistically preferable to the 2-spot model; the log-likelihood of the former being only 1.7 higher than that of the latter, i.e., a difference smaller than the uncertainties on the log-likelihood. In other words, both models are equally good descriptions of the data.}. The line "3~spot1" considers only the first and second spot and the line "3~spot2" the first and third spot to find the geometry of the off-centred dipole. The line "3~mean" interprets the last 2 hot spots as emanating from a same polar cap where the centre is approximately located in the middle of the 2 spot centres $(\Theta_p+\Theta_s)/2$ and $(\phi_p+\phi_s)/2$ (which is an approximation of the exact middle joining both centres on the sphere). With this assumption we obtain other parameters that better agree with the 2~spot model referred as ST+PST in Table~\ref{tab:NICER1}.
\begin{table*}
	\centering
	\begin{tabular}{lccccccccccccc}
		\hline
		Model & $\Theta_p$ & $\phi_p$ & $\xi_p$ & $f_p$ & $\Theta_s$ & $\phi_s$ & $\xi_s$ & $f_s$ 
		& $\zeta$ & $\alpha$ & $\beta$ & $\delta$ & $\epsilon$ \\
		 & (rad) & (cycle) & (rad) & & (rad) & (cycle) & (rad) & & (deg) & (deg) & (deg) & (deg) & ($d/R$) \\
		\hline \hline
		2 spot  & 2.251 & 0 & 0.035 & 5.347 & 2.417 & 0.459 & 0.033 & 15.490 &  49 & 85 & 58 & 171 & 0.70 \\
		3 spot1 & 2.270 & 0 & 0.036 & 5.352 & 2.417 & 0.460 & 0.033 & 15.769 &  50 & 86 & 60 & 172 & 0.70 \\
		3 spot2	&		&   &       &       & 2.988 & 0.420 & 0.056 & 1.215  &  50 & 69 & 24 & 159 & 0.88 \\
		3 mean  & 2.270 & 0 & 0.036 & 5.352 & 2.703 & 0.440 &       &        &  50 & 77 & 33 & 165 & 0.80 \\
		\hline
	\end{tabular}
	\caption{Off-centred dipole geometry deduced from the polar cap location after \cite{miller_psr_2019}.\label{tab:NICER2}}
\end{table*}

\subsection{X-ray and $\gamma$-ray time lag}

Our light-curve fitting procedure relies heavily on the phase aligned pulse profiles. The most important characteristics of these pulse profiles are the phase of the peaks in $\gamma$-ray and X-ray. For the radio pulse profile, we prefer to use the centre of the radio pulse as the reference phase.
We think indeed that the centre is more robust because it relies on a geometrical estimate not related to the flux of the radio emission that can strongly vary between components. We define the centre of the pulse to be at a phase half way between the phases of 10\% of the maximum flux of the corresponding pulse. The three light-curves for radio, thermal X-ray and $\gamma$-ray are shown in Fig.~\ref{fig:j00300451multilambda} with vertical dashed lines depicting the phase of the mid-point at 10\% maximum values for each pulse, see the values in Table~\ref{tab:phase_lag}.

The first X-ray pulse peak arrives slightly before the pulse in the radio pulse profile whereas the second X-ray peak slightly lags the second radio pulse by a phase difference of about $\Delta \phi\approx \pm 0.025$. This ordering is compatible with the off-centred dipole geometry because the order of appearance of radio and X-rays is inverted between north and south pole for simple circular shape polar caps. Even for PSR~J0030+0451, for which one pole differs significantly from a circular shape, according to \cite{riley_nicer_2019} and \cite{miller_psr_2019}, the inversion in the phase lag between both wavelengths holds. 

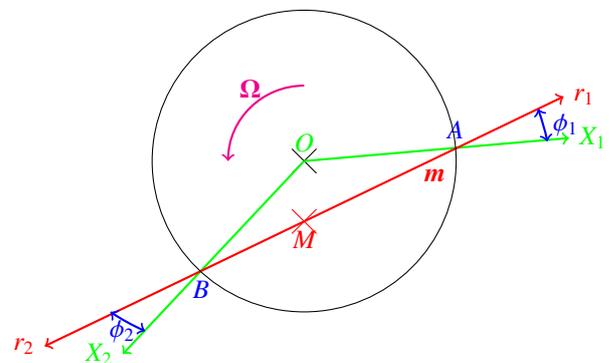
\begin{figure}
    \centering
\begin{tikzpicture}[scale=2]
\coordinate (M) at ({0, -0.4}) ;
\coordinate (m) at ({-0.853625, -0.413735}) ;
\draw (M) node[cross, red] {};
\draw[red] (M) node [below] {$M$};
\draw (0,0) node[cross] {};
\draw[blue,rotate=5] (1,0) node[above] {$A$};
\draw[blue,rotate=-133] (1,0) node[below] {$B$};
\draw[green] (0,0) node [above] {$O$};
\draw[thick,green,rotate=5,->] (0,0) -- (1.75,0) node [right] {$X_1$} ;
\draw[thick,green,rotate=-133,->] (0,0) -- (1.75,0) node [left] {$X_2$} ;
\draw[thick,red,->] (M) -- ++(1.70725, 0.82747)  node [midway,below] {$\vec{m}$} node [right] {$r_1$}  ;
\draw[thick,red,->] (M) -- ++(-1.70725, -0.82747) node [left] {$r_2$};
\draw circle(1) ;
\draw[thick,magenta,->] (0,0.5) arc (90:180:0.5) node[midway,above] {$\vec\Omega$};
\draw[thick,blue,<->]  (1.6,0.13) arc (12:20:1.6) node[midway,right] {$\phi_1$};
\draw[thick,blue,<->] (-1.275,-1) arc (-124:-114:1.5) node[left] {$\phi_2$};
\end{tikzpicture}
\caption{Geometry explaining the time lag between the radio peaks and the X-ray peaks and depicted by the angles $\phi_1$ and $\phi_2$.}
    \label{fig:time_lag}
\end{figure}

The phase lag between the radio peak and the X-ray peak is produced as follows: on one side, for a circular hot spot on the stellar surface, assuming constant and isotropic emissivity, the maximum X-ray flux is observed when the line of sight, the normal to the centre of the hot spot and the rotation axis are all coplanar. On the other side, the radio flux peaks whenever the tangent to the magnetic field line at the emission point lies in the plane defined by the rotation axis and the line of sight. Because the field lines are usually curved and not pointing in the radial direction, we expect a shift in phase between X-rays and radio. What matter is the projection of the magnetic moment and the position vectors of the hot spot centres onto the equatorial plane, the points $A$ and $B$ as shown in Fig.~\ref{fig:time_lag}. The red straight line corresponds to the projection of the magnetic moment. The two green lines depict the projection of the vector position $\vec{s}$ and $\vec{n}$. The star is rotating counter-clockwise. The first radio peak $r_1$ leads the first X-ray peak~$X_1$ by a phase shift $\phi_1$. However the second radio peak $r_2$ trails the second X-ray peak~$X_2$, separated by a phase shift $\phi_2$. The first and leading radio peak has switched to a second and trailing radio peak. This configuration always holds except when the projected magnetic axis passes by the origin~$O$. In such a case the time lag vanishes and the pulse peaks are aligned.
A last important point concerns the phase lag between radio and X-rays which is not taken into account by the thermal X-ray fitting only. The peaks in X-ray are almost aligned with the centre of the radio pulse (Table~\ref{tab:phase_lag}) with a time lag of only $0.02-0.03$ in phase. 
\begin{table}
	\centering
	\begin{tabular}{cccc}
		\hline
		wavelength & Peak~1 & Peak~2 & $\Delta \phi$ \\ 
		\hline
		$\gamma$-ray & 0.17 -- 0.18 & 0.61 -- 0.62 & 0.44  \\
		X-ray        & 0.977 & 0.500 & 0.469  \\
		radio (NRT)  & $-$0.004 & 0.473 & 0.484  \\
		$[\phi_1; \phi_2]$  $(w_{10})$ & [$-$0.088; 0.080] & [0.330;0.615] &  \\
		radio (NenuFAR)  & -0.0337 & 0.474 & 0.508  \\
		$[\phi_1; \phi_2]$  $(w_{10})$ & [$-$0.163; 0.095] & [0.330;0.619] &  \\
		\hline
	\end{tabular}
\caption{Phase location of the first and second peak in radio, X-ray and $\gamma$-ray. The phase difference~$\Delta \phi$ between both peaks is also shown (restricted to the interval [0,0.5]). The $w_{10}$ line indicates the phase interval where the radio pulses are detected above 10\% of maximal flux. Errors in phase are of the order of the size of several bins thus $\approx \pm0.01$ in X-rays and $\gamma$-rays and $\pm0.001$ in radio. \label{tab:phase_lag}}
\end{table}

\begin{figure}
    \centering
    \includegraphics[width=\columnwidth]{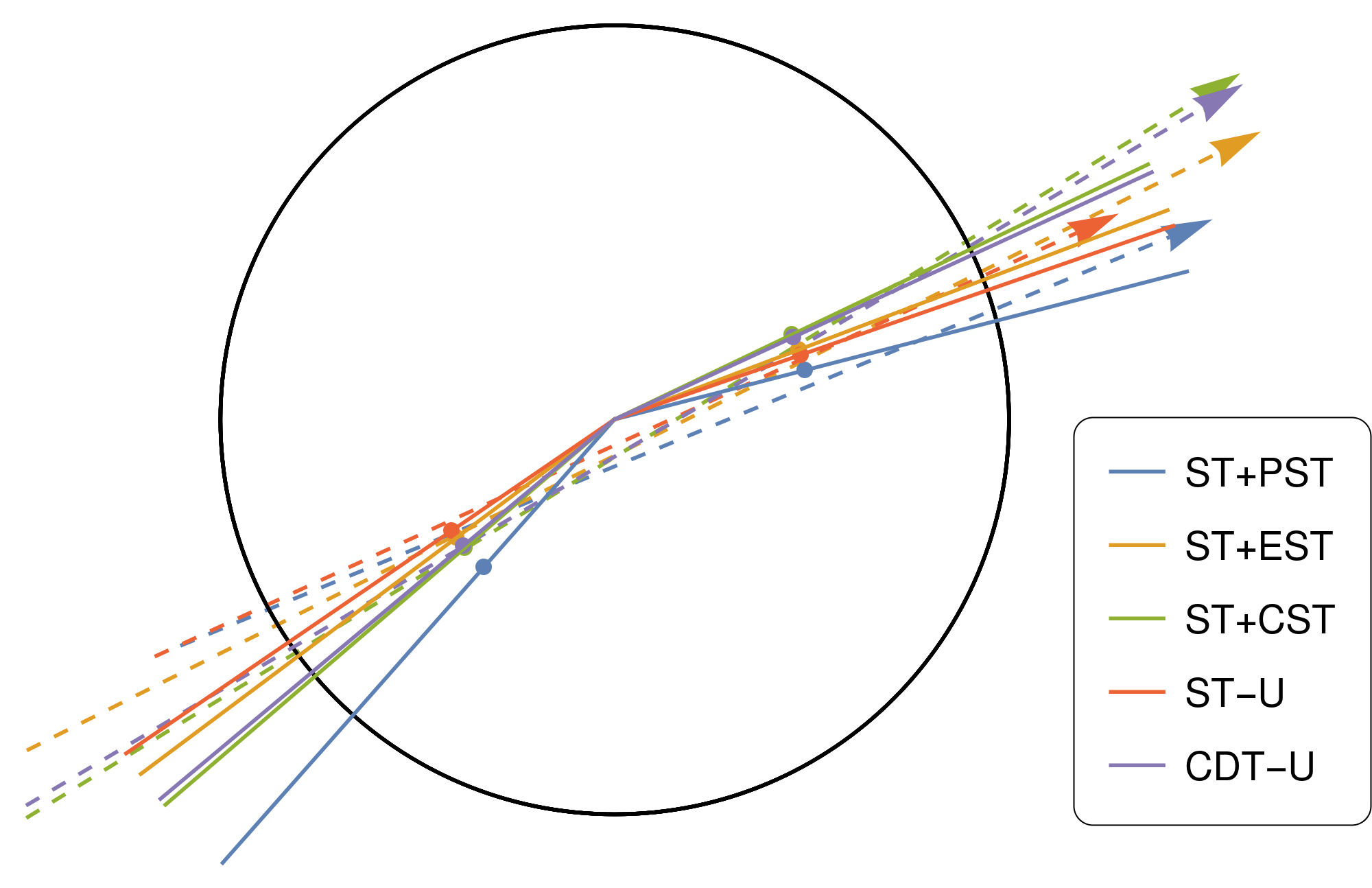}
    \caption{Time lag as expected from the NICER observations. The dashed colour lines correspond to the direction of the magnetic axis whereas the solid colour lines to the direction of the normal for each hot spot, all directions being projected onto the equatorial plane. }
    \label{fig:time_lag_NICER}
\end{figure}

Fig.~\ref{fig:time_lag_NICER} shows the different locations of the hot spots projected onto the equatorial plane as deduced from the Table~\ref{tab:NICER1}. They all give very similar directions of radio and X-ray emission phase of $\phi_1\approx6\degr$ and  $\phi_2\approx-10\degr$. Only the ST+PST model produces larger time lags of $\phi_1\approx8\degr$ and  $\phi_2\approx-26\degr$. The precise values are summarized in Table~\ref{tab:time_lag_NICER}. Note the change in sign between $\phi_1$ and $\phi_2$ because of the switch from lagging to leading peak.
The phase lags in units of the rotation period are of the order $\phi_2 \approx -0.03$ and $\phi_1 \approx 0.02$ in good agreement with the time lag in Fig.\ref{fig:j00300451multilambda}.
\begin{table}
    \centering
    \begin{tabular}{ccccc}
    \hline 
   model & $\phi_2$ (deg) & $ \phi_1$ (deg) & $\phi_2$ & $ \phi_1$ \\
    \hline
 ST+PST & $-$26 & 8 & $-$0.072 & 0.022 \\
 ST+EST & $-$10 & 6 & $-$0.028 & 0.016 \\
 ST+CST & $-$9  & 6 & $-$0.025 & 0.017 \\
 ST+U   & $-$10 & 6 & $-$0.027 & 0.015 \\
 CDT+U  & $-$9  & 6 & $-$0.026 & 0.017 \\
    \hline
    \end{tabular}
   \caption{Time lag $\phi_1$ and $\phi_2$ in degree and in units of the period, as defined in Fig.~\ref{fig:time_lag} and corresponding to the plot in Fig.~\ref{fig:time_lag_NICER}.}
    \label{tab:time_lag_NICER}
\end{table}

Accurately estimating the expected phase lag requires a more quantitative geometric study of the location of the peak X-ray emission within the hot spot as well as a careful analysis of aberration, retardation and magnetic sweep-back effects in the different wavelengths \citep{phillips_radio_1992}. Close to the neutron star, especially for the thermal X-ray emission, light-bending and the Shapiro delay should be included too. This is however out of the scope of this work.

Our multi-wavelength analysis as well as NICER thermal X-ray pulse profile fitting independently lead to a line of sight inclination angle of~$\zeta=54\degr$. As a consequence we take this value as a robust estimate of $\zeta$. Therefore from relation~\ref{eq:cos_pi_delta} we found $\alpha\approx 78\degr$ in accordance with our $\gamma$-ray fit of $75\degr$. We then take $\alpha \approx 75\degr$ as another robust result.

\subsection{Small scale surface dipole}

The two independent groups working on the thermal X-ray pulse profile modelling fitted the hot spot emission with different patches, circular shapes for \cite{riley_nicer_2019} or ovals for \cite{miller_psr_2019} who then also find possibly three hot spots. The rim of these polar caps must somehow be related to the surrounding magnetosphere, connected to the relativistic plasma flow hitting and heating the surface. It is well known that radio emission physics requires strong curvature field lines to produce sufficient electron-positron pairs and the associated radiation. This can only be achieved by small scale surface magnetic fields, dominating locally the global dipole field, possibly off-centred as discussed for instance by \cite{szary_non-dipolar_2013}. 
In this section, we show that a small amplitude surface magnetic dipole located slightly inside the neutron star crust (at a distance $0.05\,R$ below the surface) on top of a large scale off-centred dipole field can lead to polar cap shapes similar to those expected from the NICER observations.

\begin{figure}
	\centering
	\includegraphics[width=\linewidth]{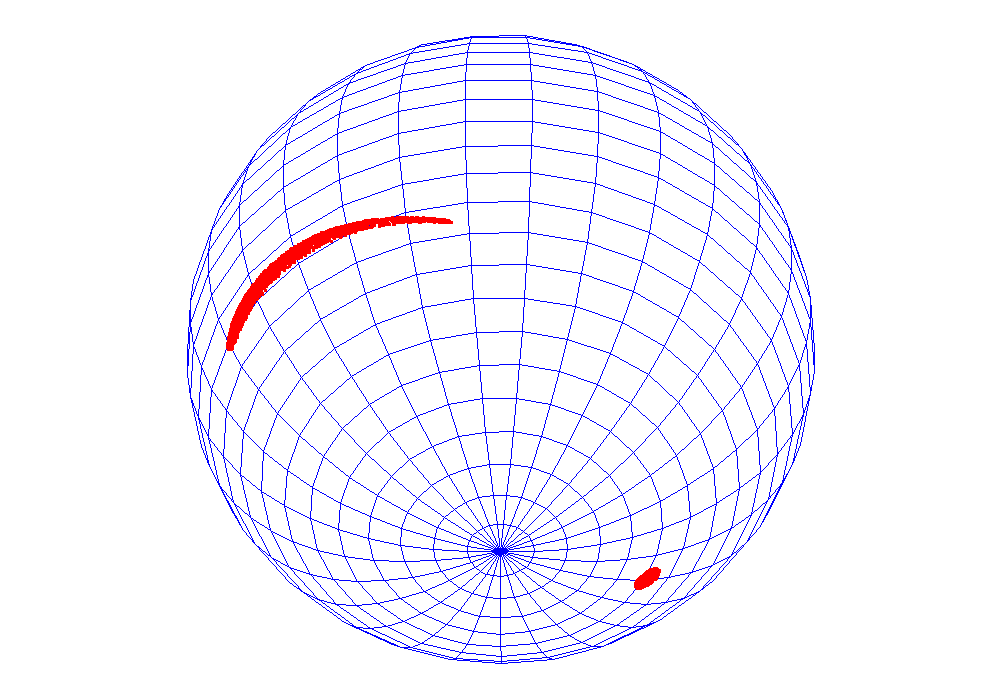}
	\caption{A 3D view of the two hot spots from the south pole. They are produced by a large scale off-centred dipole and a small scale surface dipole located close to one magnetic pole.}
	\label{fig:vue3d}
\end{figure}

There are several free parameters to specify each dipole configuration making it impossible to explore the full space parameter with high resolution and realistic magnetospheres filled with a pair plasma. Instead, here we only give a "proof of concept" of this idea by using a static vacuum dipole geometry, neglecting the displacement current. The components of the total magnetic field (off-centred dipole and small scale surface dipole) being known, we compute the field line structure by straightforward integration for a sample of representative lines. Each field line crossing the light-cylinder is reputed to be part of the open region where the plasma flows. Their foot points are connected to the hot spots. Fig.~\ref{fig:vue3d} shows an example of a two spot configuration, both poles being located in the southern hemisphere, one with a small, almost circular shape and the other much more elongated and thin, a crescent shape as reported in \cite{riley_nicer_2019}. To obtain these shapes, the parameters used in the figure are as follows: a large scale magnetic moment inclination angle of~$\alpha_0 = 75\degr$ and a position vector $(x_0,y_0,z_0) = (0,0,-0.8)\,R$, a ratio between the small scale dipole~$B_1$ and the large scale dipole~$B_0$ given by $B_1/B_0 = 0.3$, a position vector of the small dipole $(x_1,y_1,z_1) = (-0.4,-0.6,-0.6)\,R$ and an inclination angle of the small dipole amounting to $\alpha_1 = 150\degr$.
Note however that the quantitative pictures do not agree as we are not attempting to model accurately the hot spots because of our very restrictive assumptions of an empty and static magnetosphere.

\section{Discussion}
\label{sec:Discussion}

On one side, joint radio and $\gamma$-ray fitting constrains the two important angles, namely the line of sight and magnetic dipole moment inclination. On the other side, thermal X-ray pulse profiles from the surface determine the location of the hot spots and their shape. Combining both approaches in a unified framework would better constrain the priors used in the procedure of thermal X-ray fitting. Indeed, from a theoretical point of view, the heating of the stellar surface seen as hot spots must be connected to the surrounding magnetospheric activity related to pair cascades and particle bombardments. From an observational point of view, the knowledge gained from the $\gamma$-ray fitting would pin down the line of sight inclination angle and the large scale dipole obliquity, narrowing the allowed parameter space to search in the thermal X-ray pulse profile fitting.

However, a severe limitation to this approach is the number of free parameters to be added to describe the magnetic field topology. We can model the field with for instance two dipoles or a dipole plus some quadrupolar components but the number of free parameters to adjust would be similar. The thermal X-ray hot spot geometry supports the idea of an off-centred dipole locally disturbed by a small scale dipole on the surface. However, we are far from a precise quantitative picture because this would require simulating a large sample of force-free magnetospheres with dipole+dipole fields to span a reasonable range of the space parameter. In the current stage, this is computationally not feasible except if the region to search for is already small enough with only a dozen of runs to perform.

Oblateness is also neglected in our study, we assumed a perfect spherical body. According to \cite{silva_surface_2021}, the perturbation induced by the rotation of a 200~Hz neutron star like PSR~J0030+0451 is weak, the ratio between the equatorial and polar radius is expected to be between 1 and 1.04 so not too large for our qualitative study performed in this work. See for instance the vacuum and force-free oblate and prolate solutions found by \cite{petri_spheroidal_2022, petri_spheroidal_2022-1}.

Due to the presence of an interpulse component in its pulse profile, PSR~J0030+0451 is assumed to be an approximately orthogonal rotator with a line of sight close to the equatorial plane, i.e. $\zeta\sim90\degr$. But the parameters found by NICER collaboration and by our radio/$\gamma$-ray fit shows that the line of sight is closer to $\zeta = 54\degr$. This means that the radio beam cone opening angle must be large. Assuming a dipolar magnetic field structure at the radio emission site, we can estimate the emission height. Indeed the opening angle~$\theta_{\rm em}$ of the emission cone along the magnetic field lines at the surface is \citep{gangadhara_understanding_2001}:
\begin{equation}
\theta_{\rm em} \approx \frac{3}{2} \theta_{\rm pc} = \frac{3}{2} \arcsin(\sqrt{R/\rlight}) \approx 20\degr.
\end{equation}
The north pole is visible only if $\zeta \in [\alpha - \theta_{\rm em}, \alpha + \theta_{\rm em}] \approx [55\degr, 95\degr]$ which is just on the lower limit for $\zeta$. For the south pole to become visible we moreover require that $\zeta \in [\pi - \alpha - \theta_{\rm em}, \pi - \alpha + \theta_{\rm em}] \approx [85\degr, 125\degr]$ which is clearly not the case. In order to reconcile the geometry with observations, the photon emission site for radio has to be shifted to higher altitudes around $r_{\rm em} \approx 6\,R$. In this case, the new cone opening angle becomes $\theta_{\rm em}^* = 3/2 \, \arcsin(\sqrt{6\,R/\rlight}) \approx 52\degr$. Now the line of sight inclination angle must lie within $\zeta \in [\alpha - \theta_{\rm em}^*, \alpha + \theta_{\rm em}^*] \approx [22\degr, 127\degr]$ and $\zeta \in [\pi - \alpha - \theta_{\rm em}^*, \pi - \alpha + \theta_{\rm em}^*] \approx [52\degr, 157\degr]$ just on the edge to observe the south pole. This altitude would correspond to a significant fraction of the light-cylinder radius of about $6\,R/\rlight \approx1/3$. Nevertheless, the presence of a small scale surface dipole could alter the direction of radio emission of the crescent shaped pole if photons are produced relatively close to the surface, possibly shifting the emission height to lower altitudes and then seen at different phases. This could also change the radio peak separation in our model where we used a simple centred dipole in Fig.~\ref{fig:fitj00300451}.
Looking at the width of both radio pulses, they are about 110-120\degr, thus a half-opening angle of the emission cone of about $\theta_{\rm em} \sim 55-60\degr$. This agrees with the above estimate to see both pulses when $\zeta\approx54\degr$. Consequently, there is a convergence towards the fact that the radio emission emanates from high altitude, close to the light-cylinder in an almost dipolar region.

Inspecting the NenuFAR radio pulse profile in Fig.~\ref{fig:j00300451multilambda}, we observe that the emission in both radio peaks are almost phased-aligned with their NRT counterpart (see Table~\ref{tab:phase_lag}). Moreover, the second radio peak pulse width is comparable to the NRT width and located at the same phases. However, the first peak seems to depart slightly slightly from its NRT counterpart. Its width is larger and it leads the high frequency counterpart by 3\% in phase. This suggests that both radio bands are produced at approximately the same altitude within the magnetosphere. The same conclusion holds for the LOFAR observations, the width of the pulses are similar to those seen by NenuFAR although their shape differ, especially for the weaker pulse around phase $0.5$. Actually because of the small size of the light-cylinder, only $\rlight \approx 18.3\, R$, there is indeed not much space left to span a vast range in emission heights. Moreover the first peak morphology changes at low frequency, highlighting a possible change in the local magnetic field geometry at the altitude where these photons are produced and due to the small scale surface dipole. Indeed, \cite{riley_nicer_2019} points out that the first X-ray pulse profile belongs to the crescent shape hot spot. In our picture this is related to a small scale surface dipole that then also impacts the radio emission properties between the NenuFAR and NRT frequencies.

\section{Conclusions}
\label{sec:Conclusions}

With the increase in sensitivity of telescopes at all wavelengths, a simultaneous fitting of radio, X-ray and $\gamma$-ray light-curves of pulsars becomes a very powerful tool to constrain the location of the emission regions as well as the magnetic dipole obliquity and observer line of sight inclination angles. For PSR~J0030+0451, we showed that on one side, thermal X-ray pulse profile fitting and on the other side a joint radio and $\gamma$-ray light-curve fitting gave very similar expectations for those angles. This supports the idea that radio and $\gamma$-ray emission models for young pulsars apply equally well to recycled millisecond pulsars even if the presence of multipolar fields could strongly impact the surface emission and to a lesser extent the radio and $\gamma$-rays radiation. We found that $\alpha\approx 75\degr$ and $\zeta\approx54\degr$ are very robust results for the multi-wavelength light-curves of PSR~J0030+0451.

The above study checked the consistency between both approaches without digging into a precise description of the magnetic configuration. To go further would require a more quantitative analysis, performing numerical simulations of off-centred force-free dipolar magnetospheres on top of small scale surface dipoles. Unfortunately, the total number of parameters makes it intractable computationally because of the huge number of configurations to explore. It is therefore compulsory to narrow down the region of interest in the parameter space for each individual pulsar as done in the present work.

The concomitance between the two independent results in X-ray and radio/$\gamma$-ray may just be a coincidence. But with the ongoing NICER campaign, the number of pulsars with confident hot spot modelling will increase and such chance occurrence could be rejected with ever increasing confidence. Therefore, our next target is PSR~J0740+6620 \citep{miller_radius_2021, riley_nicer_2021} that shows very similar behaviour to PSR~J0030+0451. Results will be discussed in an upcoming paper.

Last but not least, millisecond pulsars might emit thermal X-rays that are polarized. The recent discussion by \cite{suleimanov_expected_2023}, based on a self-consistent atmosphere model in local thermodynamics equilibrium, shows that for an unmagnetized neutron star with a hydrogen or a carbon atmosphere, the maximum polarization degree can reach 25\% and 40\%  respectively. Moreover, it is possible for the X-ray emission to become substantially polarized if multipolar components are present, with magnetic field strength exceeding $B\sim10^{6}$~T \citep[see for example spectra and polarization of magnetized neutron stars in][]{lloyd_model_2003}.
Including polarization degree and polarization angle in the analysis could provide two additional dimensions to help to break the degeneracy in the parameter space and disentangle the surface magnetic field structure. This kind of study would be particularly relevant in the context of future X-ray polarimetry missions with sensitivity around 0.1--1~keV\footnote{Note that millisecond pulsars are too faint to be detectable by current X-ray polarimetry observatories, such as the Imaging X-ray Polarimetry Explorer mission \citep[IXPE,][]{weisskopf_imaging_2022}, which operates in the $2-8$~keV range.}.

\begin{acknowledgements}
We are grateful to the referee for helpful comments and suggestions.
This work has been supported by the CEFIPRA grant IFC/F5904-B/2018 and ANR-20-CE31-0010 (MORPHER). SG, NW and DGC acknowledge the support of the CNES. We are grateful to David Smith for helpful discussions.

The Nan\c{c}ay Radio Observatory is operated by the Paris Observatory, associated with the French Centre National de la Recherche Scientifique (CNRS). We acknowledge financial support from the ``Programme National de Cosmologie et Galaxies'' (PNCG) and ``Programme National Hautes Energies'' (PNHE) of CNRS/INSU, France. 

This paper is partially based on data obtained using the NenuFAR radio-telescope. The development of NenuFAR has been supported by personnel and funding from: Nan\c{c}ay Radio Observatory (ORN), CNRS-INSU, Observatoire de Paris-PSL, Universit\'{e} d’Orl\'{e}ans, Observatoire des Sciences de l’Univers en r\'{e}gion Centre, R\'{e}gion Centre-Val de Loire, DIM-ACAV and DIM-ACAV+ of R\'{e}gion Ile de France, Agence Nationale de la Recherche.

LOFAR, the Low-Frequency Array designed and constructed by ASTRON, has facilities in several countries, that are owned by various parties (each with their own funding sources), and that are collectively operated by the International LOFAR Telescope (ILT) foundation under a joint scientific policy. The observations
used in this work were made during ILT time allocated under the long-term pulsar monitoring campaign (projects LC0\_014, LC1\_048, LC2\_011, LC3\_029, LC4\_025, LT5\_001, LC9\_039, LT10\_014, LC14\_012, LC15\_009, LT16\_007 and LC20\_014).
We acknowledge support from Onsala Space Observatory for  the provisioning of its facilities/observational support.
The Onsala Space Observatory national research infrastructure is funded through Swedish Research Council grant No 2019-00208.

We acknowledge the use of the Nançay Data Center computing facility (CDN - Centre de Donn\'{e}es de Nançay). The CDN is hosted by the  Nan\c{c}ay Radio Observatory in partnership with Observatoire de Paris, Universit\'{e} d'Orl\'{e}ans, OSUC and the CNRS. The CDN is supported by the Region Centre Val de Loire, d\'{e}partement du Cher.

IPK acknowledges the support of Collège de France by means of “PAUSE – Solidarité Ukraine” and of NAS of Ukraine by a Grant of the NAS of Ukraine for Research Laboratories/Groups of Young Scientists of the NAS of Ukraine (2022–2023, project code "Spalakh").

\end{acknowledgements}


\end{document}